\documentstyle[epsf]{article}
\def\tilde{\widetilde}
\def\bar{\overline}

\def\*{\star}
\def\({\left(}          
\def\){\right)}         
\def\[{\left[}          
\def\]{\right]}

\def\frac#1#2{{#1 \over #2}}

\def\2pi{\hbox{$2\pi i$}}

\def\dsl{\raise.15ex\hbox{/}\kern-.57em\partial}
\def\Dsl{\,\raise.15ex\hbox{/}\mkern-.13.5mu D}
\def\ga{\gamma}         \def\Ga{\Gamma}
\def\be{\beta}           
\def\al{\alpha}

\def\la{\lambda}        \def\La{\Lambda}

\def\vphi{\varphi}
\def\CA{{\cal A}}       \def\CB{{\cal B}}       \def\CC{{\cal C}}
\def\CD{{\cal D}}              
              
              \def\CL{{\cal L}}
\def\CM{{\cal M}}       \def\CN{{\cal N}}       
       \def\CQ{{\cal Q}}       
       \def\CT{{\cal T}}

\def\debut{ \begin{eqnarray} }
\def\fin{ \end{eqnarray} }
\def\non{ \nonumber }

\def\w={={\kern -.55cm {\ }^{{\ }^{w}}}\ }

\def\presentation{
\voffset -.50in
\hoffset -.19in
\oddsidemargin 0in \evensidemargin 0in
\marginparwidth .75in \marginparsep 7pt \topmargin 0in
\headheight 12pt \headsep .25in
\footheight 18pt \footskip .35in
\textheight 9.5in \textwidth 6.5in
\columnsep 10pt \columnseprule 0pt }

\presentation

\begin{document}
\rightline{LPTHE-98-10}
\vskip 1cm
\centerline{\LARGE Quasi-classical Study of Form Factors}
\bigskip
\centerline{\LARGE in Finite Volume.}
\vskip 2cm
\centerline{\large 
Feodor A. Smirnov 
\footnote[0]{Membre du CNRS}
\footnote[1]{On leave from Steklov Mathematical Institute,
Fontanka 27, St. Petersburg, 191011, Russia} }
\vskip1cm
\centerline{  Laboratoire de Physique Th\'eorique et Hautes
Energies \footnote[2]{\it Laboratoire associ\'e au CNRS.}}
\centerline{ Universit\'e Pierre et Marie Curie, Tour 16 1$^{er}$
		\'etage, 4 place Jussieu}
\centerline{75252 Paris cedex 05-France}
 \vskip2cm
{\bf Abstract.}  We construct the quasi-classical approximation
of the form factors in finite volume using the separation of
variables. 
The latter is closely related to the Baxter equation.

\section{Introduction.}
There is an important open problem of
describing  the matrix elements
of local fields taken between two
eigenstates of Hamiltonian  (form factors ) for integrable
field theory with periodical boundary conditions.
There are at least two reasons why this problem
is interesting. First, the knowledge of
such matrix elements would allow us to study the correlators at
finite temperature.
Second, detailed understanding of the 
periodical problem would give controllable interpolation between
a massive field theory and its ultra-violet, conformal, limit.

In the present paper we shall study  purely 
conformal case, more exactly, $c<1$ models of Conformal Field Theory (CFT).
Certainly, in CFT the correlators can be found
explicitly, so one might find our study to be rather of 
academic interest. However this is not quite the case.
The point is that we study the conformal field theory in
its integrable formulation. The latter allows  deformation
to non-conformal case.

The integrable structure of CFT was
first discussed by Zamolodchikov \cite{zam}
who constructed examples of higher local integrals.
The existence of infinitely many local integrals was
proven in \cite{ff}. The spectrum of the local integrals
is a subject of
detailed study in the series of papers \cite{blz1,blz}.
In particular, the paper \cite{blz} provides
detailed investigation of famous
Baxter equation. We are trying to combine the results
of this paper with the method of separation of variables
in integrable models developed by Sklyanin \cite{skl}.
In the latter method the solutions of Baxter equations play
a role of wave functions for separated variables.
We also use intensively the relation to the classical
periodical solutions which are related to Riemann surfaces on
which the separated variables represent divisors.

The CFT compactified on the circle 
depends trivially on the
length
$L$ of the circle. However, the case $L\to\infty$ is to
be considered separately. Simple reasoning shows that
the limit of the matrix elements in this limit must
reproduce known form factors for Sine-Gordon
(more exactly restricted Sine-Gordon) model
in infinite volume \cite{book}.
Here we find a relation to the paper \cite{bbs}
where the formulae for the form factor in the infinite volume
have been explained in terms of separated variables.
It should be said, however, that in the paper \cite{bbs}
we failed to reproduce the quasi-classical limit of
form-factors completely. More careful approach of the present
paper will allow to find the missed pieces which are, in fact,
due to the contribution of "vacuum" particles.

As it will be clear from the paper, though our results are
quasi-classical, exact quantum formulae are to be found by
similar means. How efficient these formulae would be is
another question.

Finally, I would like to say that the mathematical machinery used
for study of periodical solution is the theory of Riemann surfaces.
In $L\to\infty$ case these surfaces turn into degenerate surfaces
which correspond to soliton solution. What should we achieve by studying
the quantum periodical problem is a certain deformation
of Riemann surfaces. For soliton case this deformation is explained
in \cite{def}, \cite{bbs1}; we hope to have something even more
interesting in periodical case.
\section{The formulation of the problem.}

In this paper we shall consider the $c<1$ CFT compactified on the circle
of length $L$. We have the Virasoro algebra with generators $\CL _n$
satisfying the commutation relations:
$$\[\CL_m ,\CL_n\]=(m-n)\CL_{m+n}+c\ \delta _{m,-n}{n^3-n\over 12} $$
where 
$$c=13-6\(\hbar +{1\over\hbar}\)$$
The Virasoro generators are obtained by usual construction from the
Bose field $\varphi (x)$ defined as follows
$$\varphi (x)=iQ +{2\pi ix\over L}\ P
+\sum\limits _{n\ne 0}{a _n\over n}\ e^{-{2\pi ixn\over L}}$$
where the generators 
of the Heisenberg algebra satisfy the commutation
relations:
$$[P,Q]={\hbar\over 2 i},\qquad [a_m,a_n]=
\delta _{m, -n}\ {n\hbar\over 2}$$
The field $\varphi$ is quasi-periodic:
$$\varphi (x+L)=\varphi (x)+iLP$$
This Heisenberg algebra allows the representsation with highest
vector $|p\rangle$:
$$a_n\ |p\rangle =0,\ n>0;\qquad P\ |p\rangle =p\ |p\rangle $$
later we shall use another notation for the zero-mode:
$\phi ={L\over \pi}p$.

From this Bose field we construct the operator $T(x)$ :
$$T(x)={1\over\hbar}\(:\varphi '(x)^2:+(1-\hbar)
\varphi ''(x) +{\hbar ^2\over 24}\)$$
The representation of the Virasoro algebra in the
space of representation of the Heisenberg algebra is defined
because the Fourier components of $T(x)$ satisfy the Virasoro
commutation relations:
$$T(x)=L^{-2}\(\sum\limits _{n=-\infty}^{\infty}\CL _n
\ e^{-{2\pi ixn\over L}}-{c\over 24}\)$$
obvioulsly, $T(x)$ is periodical function of $x$.

As it is shown in \cite{zam} the CFT allows integrable structure.
It means that in proper completion of the
universal enveloping algebra of Virasoro algebra one can
find a commutative subalgbra of local integrals of motion.
Namely, there exist an infinite sequence of local operators
$T_{2k}(x)$ (for $k=1,2,\cdots$) such that the operators
$$I_{2k-1}=\int\limits _{0}^LT_{2k}(x)dx $$
commute with each other. The local operators $T_{2k}(x)$
are from the module of $1$, i.e. they are costructed by taking
derivatives of $T(x)$, multipying and normal ordering them.
In particular
\debut
&T_2(x)=T(x), \qquad T_4(x)=:T^2(x):\non\\  
&T_6(x)=:T^3(x):+{c+2\over 12}:(T'(x))^2:
\non\fin
The first local integral ($I_1$) coincides with $\CL _0$.
The spectrum of $\CL _0$ is highly degenerate, but other 
local integrals of motion reduce drustically this
degeneracy.
There are two important problems. First, one has to describe this spectrum.
This problem is very complicated, but as it is conjectured in \cite{blz}
the spectrum can be defined from solutions of Baxter equations (we shall
discuss this later).

The second interesting problem consisits in calculating the matrix
elements of local operators between the eigen-states. Suppose that we
have two  eigen-states $|\Psi\rangle$ and $|\Psi '\rangle$ such that
\debut I _{2k-1}|\Psi\rangle =i _{2k-1}|\Psi\rangle ,\qquad
I _{2k-1}|\Psi'\rangle =i _{2k-1} '|\Psi '\rangle \label{mel}\fin
where $i,i'$ are eigen-values. We are interested in the matrix elements of the
kind 
$$\langle \Psi |O(x)|\Psi '\rangle $$
where $O(x)$ is certain local operator. For simplicity we can take
$O(x)$ from the module of $1$, in that case, obviously, the states
must have equal zero-modes, otherwise the matrix element vanishes.
In this paper we shall consider the matrix elements between two
states with $\phi =n$.
Obviously, it is enough to consider the matrix elements of $O(0)$
because $|\Psi\rangle$ and $|\Psi '\rangle$ are eigen-states of $\CL _0$,
so, the $x$ dependence is trivial. 

What can be said in general about the matrix elements (\ref{mel})? 
There are two
situations when they are known.
\newline
1. Consider the ``small'' states, i.e. the ones created from
Virasoro vacuum by applying few rising operators.
On these states the local integrals can be diagonalized
explicitely, and the matrix elements can be found by brut force.
\newline
2. More interesting case is the case of ``big'' states.
There are two equivalent definitions of these states.
It is a proper place to explain why we keep the lenght $L$
in all formulae. In conformal theory we can always rescale
the circle to one of length $2\pi$ passing to the
variable $y={2\pi\over L}$. The local integrals must be rescaled
as follows
$$\tilde{I}_{2k-1}=\({L\over 2\pi}\)^{2k -1}I_{2k-1}$$
where $\tilde{I}$ are the integrals on the $2\pi$-circle.
Consider the the states on $L$-circle for which the eigen-values
of the local integrals remail finite in the limit $L\to\infty$
or, equivalently the states on $2\pi$-circle for which the
eigen-values of $I_{2k-1}$ are of order $L^{2k-1}$ for certain big
parameter $L$. This is the definition of ``big'' states.
We prefer the first interpretation that is why we keep $L$ in
our formulae. It is rather clear that in $L\to\infty$ limit
the matrix elemnts between ``big'' states must
reproduce the form factors in restricted
SG theory. Indeed the large $L$ limit corresponds to
conformal thory in infinite volume. This theory can be described by
massless S-matrices \cite{zz} and form factors which don't
differ from those of massive theory. The states with $\phi =n$
correspond to $n$-soliton states in infinite volume limit.

\section{The classical periodical problem for KdV.}
Let us present several facts concerning the classical KdV hierarchy with 
periodical boundary conditions
following mostly the book \cite{nov}. We have the field $u(x)$ satisfying the
periodicity conditions
$u(x+L)=u(x)$, the second (Magri) Poisson
structure is defined by
\debut
\{u(x_1),u(x_2)\}= (u(x_1)+u(x_2))\delta '(x_1-x_2) +
\epsilon  \delta '''(x_1-x_2)
\label{poisson}
\fin
We shall consider two "real forms": the case $\epsilon =+$
corresponding to usual real solutions of KdV equation
(this case is denoted by rKdV), and the case $\epsilon =-$
which corresponds to certain class of complex solutions
(this case will be denoted by cKdV). 
The latter case is related to
the $c<1$ models of CFT
discussed in the previous section because in the quasi-classical
limit the Poisson brackets of $u(x)=\hbar T(x)$ coincide with 
(\ref{poisson}). 
Actually the two cases must be understood as analytical
continuations of each other, see \cite{bbs} for more expanations.

The KdV hierarchy possesses infinitely many integrals
of motion $I_{2k-1}$ ($k\ge 1$) the first of them
(the momentum) being
$$I_1=\int\limits _{0}^L u(x)\ dx $$
Making
the rescaling
$$y={2\pi\over L}x,\qquad \tilde{u}(y)=L^2u\({L\over 2\pi }y\)$$
we map the periodical problem for $u$ with arbitrary $L$ to the
one for $\tilde{u}$ with the period equal $2\pi$, the integrals of motion
scale as in the quantum case.

The exact solution of KdV equation is due to existence of
Lax representation. The auxiliary linear problem is
\debut
\( {d^2\over dx^2} -u(x)\)\psi (x,\la)=\la ^2 \psi (x,\la)
\label{alp} \fin
It is convenient to rewrite this equation as matrix first order equation.
To this end introduce the field $\vphi (x)$ related to $u(x)$ by
Miura transformation:
$u=(\vphi ')^2+\vphi ''$.
The field $\vphi $ is real for rKdV and pure imaginary for cKdV.
The equation (\ref{alp}) is equivalent to
the following one:
$${d\over dx}\Psi(x,\la)=\CL (x,\la)\Psi(x,\la)$$
where
\debut
\Psi(x,\la)=\left(
\matrix{e^{-{\vphi (x)\over 2}}\psi(x,\la)\cr
e^{{\vphi (x)\over 2}}\(\psi(x,\la)'-\psi(x,\la)\vphi (x)'\)}
 \right) \non
\fin
and
\debut
\CL (x,\la) =\pmatrix{
{\vphi(x)'\over 2},& e^{-\vphi (x)}\cr
\la ^2 e^{\vphi (x)},&-{\vphi(x)'\over 2}}\non\fin
The fundamental role is played by the monodromy matrix:
$$\CT _{x_0}(\la)=P\exp \(\int\limits _{x_0}^{x_0+L}\CL(x,\la)dx\)$$
The trace of $\CT _{x_0}(\la)$ which does not depend on $x_0$ is
denoted by $T(\la)$. Let us recall the main properties of $T(\la)$.
The function $T(\la)$ is an entire function of $\la ^2$ with infinitely
many zeros accumulating to $\infty$ along the negative real axis.
The simplest example is given by $u=0$ for which
$$T(\la)=2\cos (L\sqrt{-\la ^2})$$
The graph of this function looks as follows:

\hskip 2cm\epsffile{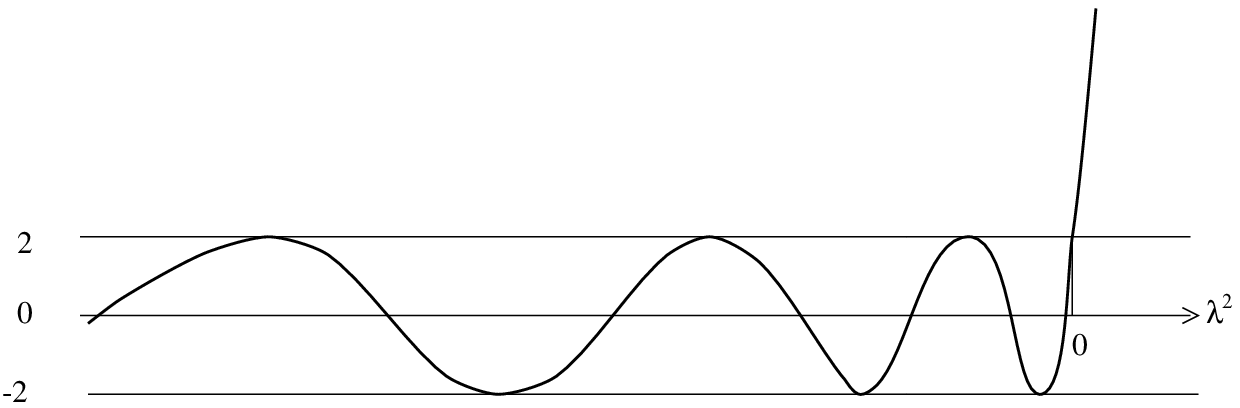}
\vskip 0.3cm
{\it Figure 1}
\vskip 0.3cm
Further we shall give other examples.
The monodromy matrix is a $2\times 2$ matrix with unit determinant, so,
$$T(\la)=\Lambda(\la)+\Lambda(\la)^{-1}$$
where $\Lambda$, $\Lambda ^{-1}$ are the eigenvalues of $\CT$.
It is important to notice that $\Lambda (\la)$ is not an entire
function of $\la$, it has quadratic branch points at zeros
of odd order of the discriminant $\Delta(\la ^2)=T(\la)^2-4$.
The discriminant is an entire function of $\la ^2$
with asymptotic following from (\ref{as}), so,
it can be described by
converging infinite product over its zeros
$$\Delta(\la ^2)=C
\ \la ^2\prod\limits \(1-{\la^2\over \nu _i ^2}\)^{k_i}$$
where $C=d\Delta/d\la ^2(0)$, $k_j=1,2$.
The asymptotical behaviour of $\La (\la)$ is governed by the local integrals of
motion:
\debut
\log \La (\la)\sim \la L+\sum\limits _{k\ge 1}\la ^{-2k+1}I_{2k-1}\non\\
\la \to\infty,\qquad {\rm Re} \la >0\label{as}\fin

The function $\Lambda (\la)$ is a singe-valued function on the
hyper-elliptic Riemann surface $\Sigma$ (generally of infinite genus)
given by the equation
$$\mu^2 =\la ^2\prod\limits _{k_i=1}(\la ^2 -\nu_i^2)$$
The tractable mathematically and (fortunately) at the same time most
interesting physically case is when $\Sigma$ has finite genus, i.e.
when $\Delta(\la ^2)$ has only finitely many zeros of first order.
The typical situation of this kind is given by
the periodical analogues of $n$-soliton solutions.
It that case we have
simple zero at $\la =0$,
$2n$ real positive single zeros of $\Delta$
(we denote them by $\la _1 ^2 ,\cdots ,\la _{2n} ^2$ and infinitely
many negative double zeros (we denote them by $-\mu _{-1}^2,-\mu _{-2}^2
\cdots$).
The graph of the function $T(\la)$ looks as follows:

\vskip 1 cm\hskip 2cm\epsffile{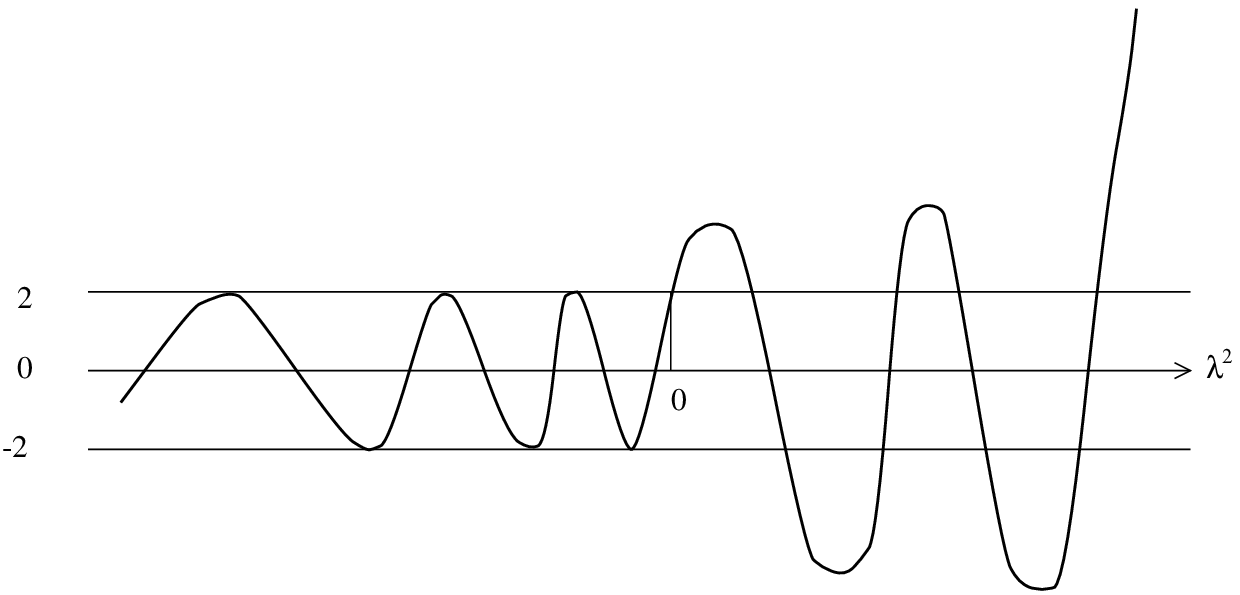}
\vskip 0.3cm
{\it Figure 2}
\vskip 0.3cm
The segments of the real axis of $\la ^2$ where $|T(\la )|>2$
are called forbidden zones of the periodic potential $u$
(there are no bounded wave-functions for these values of energy).

The hyper-elliptic Riemann 
surface $\Sigma$ of genus $n$ is described by
$$\mu^2=\la ^2 P(\la ),\qquad P(\la )=\prod\limits _{i=1}^{2n}
(\la ^2-\la _i^2)$$
Conventionally it is realized as two-sheet covering of the $\la ^2$
plane with cuts and canonical basis of homology chosen as
follows

\hskip 1.5cm\epsffile{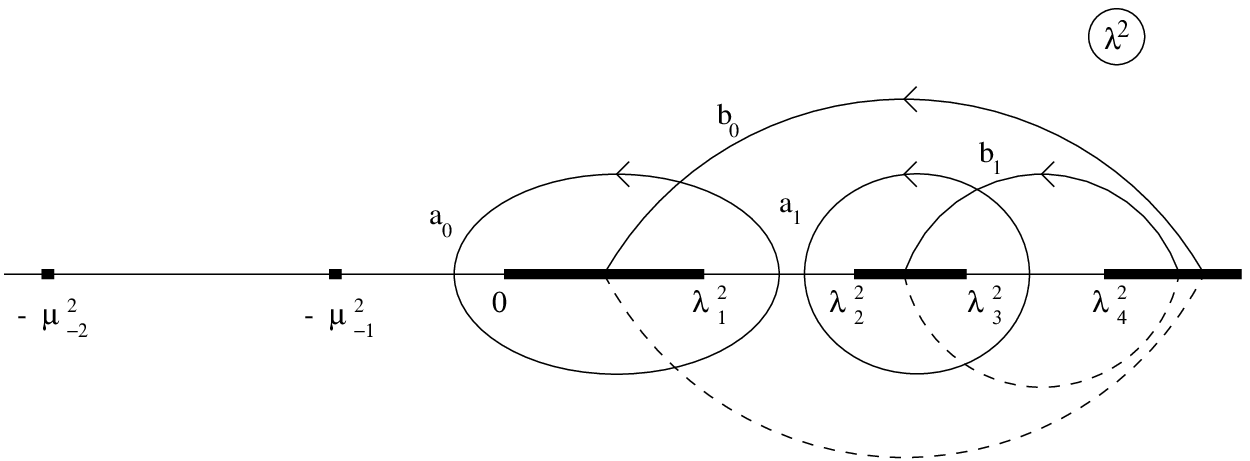}
\vskip 0.3cm
{\it Figure 3}
\vskip 0.3cm
We shall be mostly using another model of this surface
considering it as $\la$-plane with cuts:

\hskip 3.5 cm\epsffile{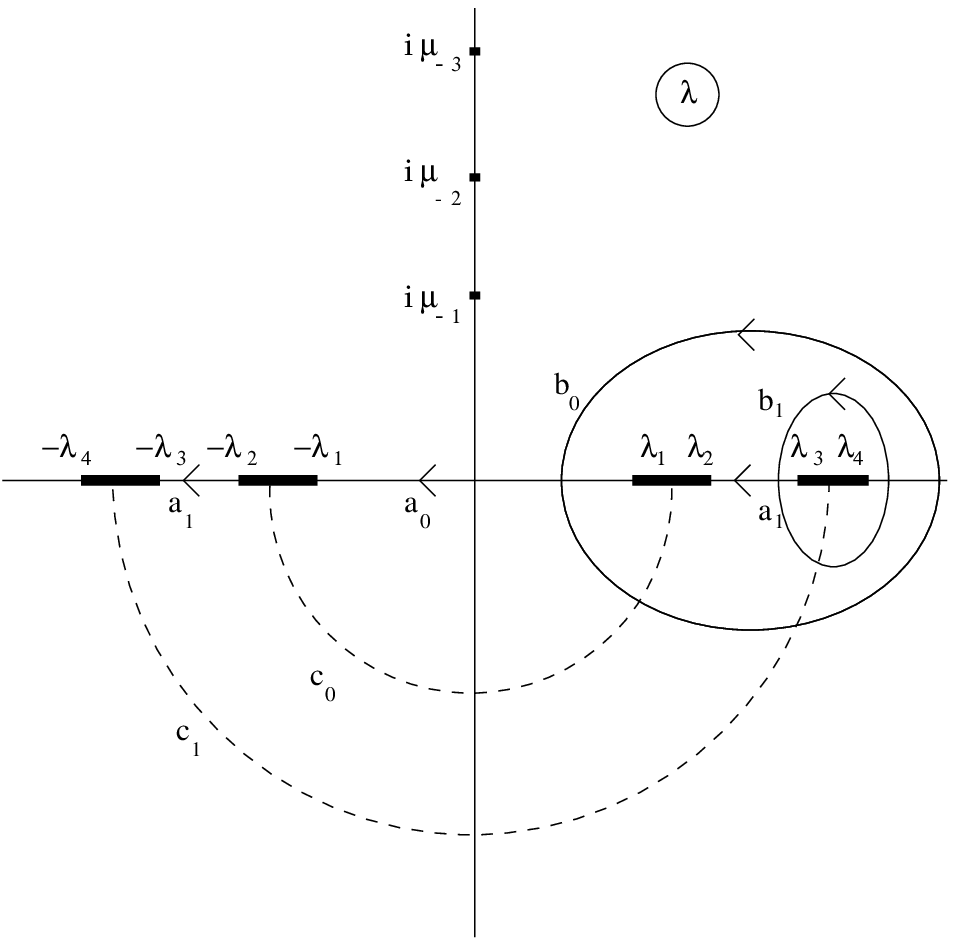}
\vskip 0.3cm
{\it Figure 4}
\vskip 0.3cm
The upper (lower) bank of the cut $[\la _{2j-1},\la _{2j}]$ is identified
with the upper (lower bank of the cut $[-\la _{2j-1},-\la _{2j}]$.
Obviously the upper (lower) half planes correspond to
first (second) sheet of the surface in usual realization.

The function
$\Lambda (\la) $ continues to the lower half-plane as
$$\Lambda (-\la)=\Lambda (\la)^{-1}$$
Let us consider in some details properties of this
function.

The function $\log \Lambda (\la )$ is called quasi-momentum
because the equation
(\ref{alp}) has the Floquet solution $\psi (x, \la)$ which
satisfies quasi-periodicity condition:
$$\psi (x+L, \la) =  \La (\la ) \psi (x, \la) $$
another name for this solution is Baker-Akhiezer (BA) function.
The BA function is single-valued function on $\Sigma$,
$\psi (x, \la)$ and $\psi (x, -\la)$ give two linearly independent
for generic $\la$ quasi-periodical solutions of (\ref{alp}),
the second one corresponds to the quasi-momentum $-\log \La (\la)$.

The function $\Lambda (\la)$ is single-valued 
on $\Sigma $ with essential
singularity at infinity. Hence $d\log \La(\la)$ is abelian differential
on $\Sigma$ with second order pole at infinity:
$$ d\log \La (\la)=(L+O(\la ^{-1}))d\la $$
(the local parameter at $\infty$ is $\la ^{-1}$).
Moreover it must be normalized:
\debut
\int\limits _{a_j}d\log \La =0\quad \forall j  \label{norm}
\fin
because as it is seen on fig.2 $\La(\la)$ is real positive
or negative function in the interval corresponding to $a$-cycle.
Then in order that $\Lambda (\la)$ is single-valued we need
that
\debut
\int\limits _{b_j}d\log \La = 2\pi i k_j
\label{period}
\fin
for some integer $k_j$. This requirement can not be satisfied
for  arbitrary $\Sigma$, it is actually a restriction on the moduli
of the surface. For the periodical analogues of $n$-soliton solutions
we have $k_j=n-j+1$
which corresponds to the fact that exactly $n-j+1$ simple zeros of $T(\la)$
are surrounded by $b_j$. This is exactly the situation
presented on the fig.2, in more general case between two forbidden zones
$T(\la)$ can make several oscillations from -2 to 2.
Provided (\ref{norm}) and (\ref{period})
are satisfied $\log \La (\la)$ is a function defined on $\Sigma $ with
cuts along the $a$-cycles whose jumps on $a_j$ equal $2\pi i k_j$.

The dynamics of finite-zone solution is conveniently described by
motion of zeros of BA function. There are exactly $n$ of them
($\ga _0,\cdots ,\ga _{n-1}$). Let us present for completeness the equation
describing the $x$-dependence of $\ga _j$:
\debut
{\partial\over \partial x} \ga _j ={\sqrt{P(\ga _j)}\over\prod _{k\ne j}
(\ga _j^2-\ga _k^2)}
\fin
The dynamics with respect to higher times is described by similar
equations.
The dynamics is linearized by Abel
transformation of the divisor of zeros of BA function onto
Jacobi variety of $\Sigma$. Recall that we consider two different
real forms of KdV
(rKdV and cKdV). The points of divisor corresponding to this
two real forms move along topologically equivalent, but
geometrically different trajectories.
In rKdV case $\ga _j$ moves along the cycle $a _j$ as it is
drawn on the fig.4.
In cKdV $\ga _i$ runs along a trajectory close to the cycle $c _j$
on the fig.4. Clearly, the half-basis of $c$-cycles
is equivalent to the half-basis of $a$-cycles.

\section{Hamiltonian structure of finite-zone solutions.}

Let us discuss the most subtle issue in the theory of
finite zone integration, namely, the Hamiltonian description
of the solutions. The surface $\Sigma $ is parametrized by $2n$
parameters ($\la _1,\cdots ,\la _{2n}$). These parameters are not
all independent, they are subject to $n$ restrictions (\ref{period}).
So, we are left with $n$-dimensional sub-manifold ($\CM$) in the moduli
space of hyper-elliptic surfaces. It is convenient to parametrize $\CM$
by $\tau _1,\cdots ,\tau _n$ such that $\tau _i ^2$ are positive
zeros ot $T(\la)$ in $\la ^2$-plane. Earlier we have introduced
the variables $\ga _0,\cdots ,\ga _{n-1}$. The phase space of finite-zone
solution is the $2n$-dimensional manifold locally described
by the coordinates $\{\tau _1,\cdots ,\tau _n,\ga _0,\cdots ,\ga _{n-1}\}$
is embedded into the infinite-dimensional phase space
of KdV. Restricting the symplectic form which corresponds to Poisson
structure (\ref{poisson}) to this finite-dimensional manifold
we obtain the symplectic form $\omega =d\alpha$ with 1-form $\omega$
given by
\debut
\alpha =\sum\limits _{j=0}^{n-1} \log \La (\ga _j){d \ga _j  \over \ga _j }
\non\fin

Our main concern is quantization, so, we have to ask ourselves
the question whether the quantization of this
finite-dimensional mechanics is relevant to the real quantization
of KdV. Logically, the answer to this question is negative
because restricting ourselves to the finite-dimensional
sub-manifold we ignore a good deal of quantum fluctuations allowed in the
infinite-dimensional phase space. However, as is shown in \cite{bbs1}
for the case of solitons ($L\to \infty$ limit) the quantization
of the finite-dimensional system gives an important piece of
exact quantum answer for the matrix elements (form factors \cite{book}).
To understand why it works and how to generalize the results of
\cite{bbs1} to periodical case we have to consider the
hamiltonian structure in some more details. This
consideration will allow also to reproduce
quasi-classically an important part of
solitons form factors which we could not do in \cite{bbs1}.

Let us analyze more general situation of which KdV provides a
particular case. Take a class of classical integrable models with
trigonometric R-matrix.  For any such system
(continuous or lattice one) the monodromy matrix
$$\CT(\la)=
\pmatrix{\CA(\la),&\CB(\la)\cr \CC(\la),& \CD(\la)}$$
can
be introduced which satisfies the famous Sklyanin's relations
\debut
\{\CT (\la)\ ,\hskip -0.2 cm {}^{\otimes}\ \CT(\mu)\}=\[r(\la ,\mu),
\CT (\la)\otimes \CT(\mu)\]
\label{rttcl}\fin
with trigonometric R-matrix:
$$r(\la ,\mu) ={1\over \la ^2-\mu ^2}
\pmatrix{\la ^2+\mu ^2, &0,&0,&0\cr
0,&0,&2\mu ^2,&0\cr0,&2\la ^2,&0,&0\cr 0,&0,&0,&\la ^2+\mu ^2}$$
For a lattice regularization of KdV \cite{fv} the monodromy
matrix is a polynomial in $\la ^2$ of degree $N$. The determinant $D(\la )$ of
$\CT(\la)$ is in the center of the Poisson algebra. The trace $T(\la )$
of $\CT(\la)$ is a generating function of $N$ independent
integrals of motion. To describe the phase space one has to
introduce $N$ coordinate-like variables. Following general
approach developed in \cite{fml,skl} we take as such zeros of
$\CB (\la)$:
$$ \CB (\la)=\CB (0)\prod\limits _{j=1}^N \(1 -\({\la\over\ga _j}\)^2\)$$
Using (\ref{rtt}) one finds:
\debut
\{\La (\ga _i),\La (\ga _j)\} =\{\ga _i,\ga _j\} =0\qquad
\{\La (\ga _i),\ga _j\}= \delta _{i,j}\ga _i\La (\ga _i)\non
\fin
where $\La (\la)$ is eigenvalue of $\CT (\la )$:
$$ \La (\la) ={T(\la)+\sqrt{\Delta (\la )}\over 2}$$
the discriminant $\Delta (\la ) =T(\la)^2-4D(\la)$.
Thus the variables $\log {\ga _j}$ and $\log \La ({\ga _j}) $
are canonically conjugated and the symplectic form can be
written as $\omega =d\alpha $ with
\debut
\alpha =\sum\limits _{j=1}^N \log \La (\ga _j) {d\ga _j\over \ga _j}
\label{simf}
\fin
Generally, the dynamics is linearized on the Jacobi variety
of the
hyper-elliptic Riemann surface $\Sigma$ described by
$\mu ^2 = \Delta (\la )$. The genus of $\Sigma$
equals $N$, the points of the divisor $\ga _1,\cdots ,\ga _N $  move
along certain closed curves topologically equivalent to a
half-basis of $a$-cycles on $\Sigma$. 

Later we shall need also the Liouville measure corresponding
to this symplectic form. Taking zeros of $T(\la)$ ($\tau _1, \cdots ,
\tau _N$) and $\ga _1,\cdots ,\ga _N $ as coordinates on the phase
space one finds
\debut \wedge ^N \omega=\prod\limits _{j=1}^N{1\over
\La (\ga _j)-\La ^{-1}(\ga _j)}
\prod\limits _{i<j}(\ga _i^2-\ga _j^2)
\prod\limits _{i<j}(\tau _i^2-\tau _j^2){d\ga _1\over\ga _1}
\wedge\cdots\wedge{d\ga _N\over\ga _N}\wedge d\tau _1^2
\wedge\cdots\wedge d\tau _N^2
\label{mes}
\fin


What happens to all that in the continuous limit
which corresponds to finite-zone solution of KdV?
The degree of $\CT (\la)$ goes to infinity, but the surface $\Sigma$
turn into a surface of infinite genus of rather special type.
Namely, only finitely many of zeros of $\Delta (\la ^2)$
($0,\la _1^2, \cdots ,\la _{2n}^2$) remain
simple ones  while infinitely many zeros  ($-\mu _{-1} ^2,
-\mu _{-2}^2,\cdots$)
become double zeros. The polynomial $T(\la )$ looks
on the lattice as

\vskip 1 cm\hskip 2cm\epsffile{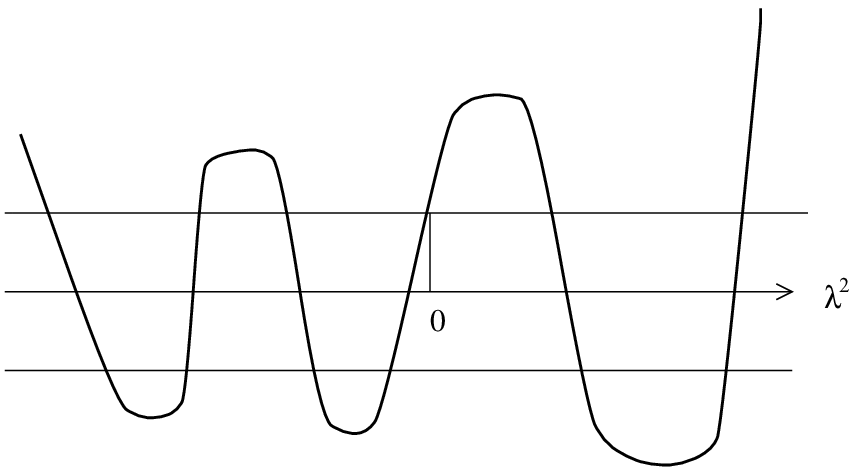}
\vskip 0.3cm
{\it Figure 5}
\vskip 0.3cm

In the continuous limit it turns into the one presented on the fig.2.
In the lattice regularization of rKdV case the points $\ga _j$ are
moving inside the zones where $|T(\la )|>2$. In $N\to\infty $
limit only $n$ of them ($\ga _0,\cdots ,\ga _{n-1} $) move inside
the $|T(\la )|>2$ zones on $\la ^2>0$ half-axis on fig.2, while
infinitely many  (we shall denote them by
$\ga _{-1},\ga _{-2},\cdots $) happen to be
confined at the points $-\mu _{-i}^2$. This is how the restriction
of degrees of freedom takes place in the classical case.
It must be emphasized that this is a dynamical procedure
which can not be carried out on quantum level. The points
$\ga _{-1},\ga _{-2},\cdots $  can not be kept at
fixed positions, but must be distributed with certain wave function
localized around their classical values.

To finish this section let us present some explicit formulae
concerning zeros of $\CB (\la)$ in the continuous case.
Consider the equation (\ref{alp}) with finite-zone
potential $u(x)$. Let us fix the normalization of
the BA $\psi (x,\la )$. If we require  that
$$\psi (\la) \sim \exp (\la x) ,\qquad \la\to\infty $$
the BA function must have $n$ simple poles in the finite
part of the plane. It is convenient to put this poles to
$n$ branch points, say, $\la _1,\cdots ,\la _n$.
We can introduce another function $\psi ^{\dag}(x,\la)$
which satisfies
$$ \psi ^{\dag}(x+L,\la) =\La ^{-1}(\la)\psi ^{\dag}(x,\la)
,\qquad \psi ^{\dag}(\la) \sim \exp (-\la x) ,\qquad \la\to\infty  $$
and has poles at the complimentary set of the branch points.
These two solutions satisfy the relations
\debut
W(\psi (\la),\psi ^{\dag} (\la))=2\la,\qquad
\psi (\la)\psi ^{\dag} (\la) ={\prod\limits _{j=0}^{n-1}(\la ^2-\ga _j^2)
\over \sqrt{P(\la )}}
\non \fin
where $W(\psi ,\psi ^{\dag} )$ is the Wronskian determinant.
From these two solutions to (\ref{alp}) one easily
reconstructs the monodromy matrix finding in particular
that
$$ \CB (\la)=(\La (\la)-\La ^{-1}(\la))\ \psi (\la)\psi ^{\dag} (\la)=\CB (0)
\prod _{j=0}^{n-1}
\(1-{\la ^2\over\ga _j ^2}\)\prod _{j=1}^{\infty}
\(1+{\la ^2 \over\mu ^2_{-j}}\)$$
Thus we find that $\CB(\la)$ has indeed zeros not only at the
moving points of the divisor, but also at the confined points $\mu _{-j}$.

\section{ Separation of variables. Baxter equation.}

Our nearest goal is to write down the quasi-classical expression for the
wave function corresponding to the periodical analogue of $n$-soliton
solution. As we have seen there are two types of coordinate-like
variables: $\ga _1,\cdots ,\ga _n$
which moves classically along the $a$-cycles and
 $\ga _{-1}, \ga _{-2},\cdots$ which are classically confined.
The first type of variables does not pose a problem for writing the
quasi-classical wave-function. The contribution to the wave-function
from the second type of variables is similar to that of a
number of harmonic oscillators in the ground state. To understand
this contribution we will need some pieces of exact quantum
information.

Consider
an operator-valued monodromy matrix $\CT(\la)$ with the same
notations for the matrix elements as before which satisfies
the quantum analogue of the Poisson brackets (\ref{rttcl}):
\debut
R(\la ,\mu)(\CT(\la)\otimes I)(I\otimes \CT(\mu))=
(I\otimes \CT(\mu))(\CT(\la)\otimes I)R(\la , \mu)
\label{rtt}
\fin
where
$$R(\la ,\mu) =\pmatrix{\la ^{2} q-\mu ^{2}q^{-1}
&0,&0,&0\cr
0,& \la ^2 -\mu ^{2},&\mu ^{2}(q-q^{-1}),&0\cr
0,&\la ^{2} (q-q^{-1}),& \la ^{2} -\mu ^{2},&0\cr
0,&0,&0,
&\la ^{2}q-\mu ^{2}q^{-1}}$$
where
$$q=e^{i\pi\hbar},$$
$\hbar$ is the Plank constant.
In conventional notations coming from SG theory \cite{book}
$$\hbar ={\xi \over \pi +\xi}$$
The quantum determinant of $\CT (\la) $ is in the center of the
algebra, in quantum KdV case it is fixed to be 1:
$$\CA(\la  )\CD(q\la )-
\CB(\la )\CC(q\la )=1$$
where $q=\exp (i\xi)$.
The monodromy matrix $\CT (\la )$ is an entire function of
$\la ^{2}$. The trace of the monodromy matrix $T(\la)$
is a generating function of integrals of motion.
In quantum KdV case one can construct the monodromy matrix
directly in continous theory \cite{blz1} by proper normal ordering of the
classical formula.

Let us review the method of separation of variables which
was developed by Sklyanin \cite{skl} who combined the general approach of
Inverse Scattering Method \cite{fst} with the ideas of \cite{gw} and
\cite{fml}. In application to our particular case the results of
Sklyanin can be formulated as follows. Consider the element $\CB (\la)$
of the monodromy matrix. It defines a commutative family of
operators due to the fact that
$$[\CB (\la ),\CB (\mu)]=0$$
Moreover $\CB (\la)$ is an entire function of $\la ^{2}$ which
grows not faster than $\exp (\la ^p L)$ 
(with certai $p$, see below) at the infinity. that is why it
can be presented as an infinite product over its zeros:
\debut \CB (\la) =
\CB(0)\prod\limits _{j=1}^{\infty}\(1-\({\la\over\ga _j }\)^{2}\)
\label{ad}\fin
The idea of Sklyanin is to consider the system in ``$\ga$-representation.

The functions $\CA(\la)$ and $\CD (\la)$ are entire functions of
$\la ^{2}$ and as such they can be expanded into series
of infinite radius of convergence:
$$\CA (\la)=\sum\limits _{n=0}^{\infty} \la ^{2n}\CA _n,\qquad
\CD (\la)=\sum\limits _{n=0}^{\infty} \la ^{2n}\CD _n $$
Following \cite{skl} one introduces the operators
$$\Lambda _j =\CA (\ga _j), \qquad \tilde{\Lambda} _j =\CD (\ga _j)$$
where the operator $\ga _j$ is substituted into $\CA$ and $\CD$
from the left (exactly replacing $\la$ in (\ref{ad})).
The operators
$\ga$, $\La$, $\tilde{\La}$ posses the following important properties:
\debut
[\ga _j,\ga _k]=0,\qquad [\La _j,\La _k]=0, \qquad
\La _j \ga _k =q^{\delta _{j,k}}\ga _k \La _j,\qquad
\Lambda _j\tilde{\Lambda} _j =1
\non\fin
first three equalities follow directly from the commutation
relations while the last one is the consequence of the
commutation relations together with the quantum determinant.
These commutation relations show that
$$\Lambda _j=\epsilon _j
\exp \(i\pi\hbar\ga _j{\partial\over\partial \ga_j}\),
\qquad
\tilde{\Lambda} _j=
\epsilon _j \exp \(-i\pi\hbar\ga _j{\partial\over\partial \ga_j}\)
$$
where $\epsilon_j =\pm 1$, the necessity of introducing
$\epsilon _j$ comes from consideration of the fig.2:
the variables $\gamma $ move (or rest) classically inside zones where
$|T(\la)|\ge 2$, so $\epsilon_j =\pm 1$ depending on whether
$T(\la) \ge 2$ or $T(\la)\le -2$.

Our goal is to diagonalize the Hamiltonians (the operator $\CT (\la)$)
in the $\ga$-representation.
This can be done if we accept the following:\newpage
{\bf Conjecture.}\newline
{\it The function }$\CB (\la) ^{-1}T(\la)$ {\it is a meromorphic function
of} $\la ^{2}$ {\it with infinitely many poles which can be expanded
in convergent series:}
$$\CB (\la) ^{-1}T(\la)=\sum\limits _{j=-\infty}^{n-1}
{1\over \la ^{2}-\ga _j ^{2}}(\Lambda _j +\tilde{\Lambda}_j)$$
\vskip 0.5cm

Assuming that
the wave-function
is presented as the infinite product:
\debut
\Psi =\prod\limits _{j=-\infty}^{n-1}Q _j(\ga _j)
\label{psi}
\fin
where every $Q_j$ satisfies the equation:
$$\epsilon _jt(\la )Q_j(\la)=Q_j(q\la)+Q_j(q^{-1}\la)$$
The sign $\epsilon _j$ can be taken away multiplying
$Q _j$ by appropriate power of $\la$, and basically we have to study the
famous Baxter's equation:
\debut
t(\la)\CQ(\la)=\CQ(q\la)+\CQ(q^{-1}\la)
\label{q+q}
\fin
where $t(\la) $ is the eigenvalue corresponding to $\Psi$:
$$T(\la)\Psi=t(\la)\Psi $$
All this procedure is called the separation of variable because
it allows to transform the infinite-dimensional spectral problem
to one-dimensional ones. The separation of variables in quantum
case gives an exact quantum analogue
of the classical separation of variables which is obvious in the formula
(\ref{simf}).

The equation (\ref{q+q}) can be considered as a finite-difference
analogue of Schr\"odenger equation, this analogy, however, is
not quite straightforward. It the case of usual Schr\"odenger
equation the wave-function belongs to certain functional space like $L_2$.
In our case the wave-function $\CQ(\la)$ is characterizes by
its analytical properties.

As a second order difference equation (\ref{q+q}) must have two solutions
up to multiplication by quasi-constant (a function of $\la ^{2\over \hbar}$).
Following \cite{blz} we require that one of them
(denoted later by $\CQ(\la )$) is
an entire function
of $\la ^{2}$.
Recall that $t(\la)$ is also
supposed to be an
entire function of $\la ^2$. The next important piece of information
is the asymptotic behaviour of these functions.
The equation (\ref{q+q}) can be rewritten as
$$t(\la)=\La _q (q^{1\over 2}\la)+\La _q ^{-1} (q^{-{1\over 2}}\la)$$
where
\debut \La _q(\la)={\CQ(q^{1\over 2}\la)\over\CQ(q^{-{1\over 2}}\la)}
\label{Laq}
\fin
is the quantum analogue of the eigenvalue of monodromy matrix.
In quantum
KdV theory $\La _q (\la)$ allows the following asymptotical expansion
around infinity:
\debut
\log \La _q(\la)\sim \la ^p L+i\phi\pi +
\sum\limits _{k\ge 1}\la ^{-(2k+1)p}I_{2k-1}\non\\
\la \to\infty,\qquad
\pi\hbar <{\rm arg}\la ^2 <\pi (1-\hbar)
\label{ass}\fin
where
$$ p={1 \over 1-\hbar}\ ,$$
$I_{2k-1}$ are quantum local integrals of motion (see \cite{blz}
for exact normalization of them). The fact that $T(\la)$ is an entire
function of $\la ^{2}$ that
has the asymptotical expansion in terms of $\la ^p$ 
is well known \cite{fst}.
The explanation of this fact is due to the renormalization of
mass. More modern and clear explanation in terms of CFT is
given in \cite{blz}. Notice that our  definition of $\la $
coincides with the one used in \cite{blz}. The phase $\phi $ equals to
the topological charge, for the periodic analogue of $n$-soliton
state $\phi = n$.
The leading terms
of the asymptotics of
$\CQ (\la)$ are supposed to be:
\debut
\log \CQ(\la)\sim {1\over 2i\sin {\xi\over 2}}\la ^p L-
{1\over \hbar}\phi\ \log \la +O(\la ^{-p}),
\qquad
\la ^2 \to\infty,\qquad 0 < {\rm arg }\la ^2 <2\pi
\label{asss}\fin
In the paper \cite{blz} one finds detailed description of complete
asymptotical series for $\CQ$.

Furthermore it is required that the function $t(\la)$ has only finitely many
zeros away from the axis $\la ^{2}<0$
($n$ real positive zeros for the periodic analogue
of $n$-soliton state) while the
function $\CQ(\la)$ has only finitely many zeros away from
the axis $\la ^{2}>0$
(for the periodic analogue
of $n$-soliton state they are absent).
The basic conjecture accepted
in \cite{blz} is that counting the solutions of (\ref{q+q})
one counts the vectors in the space of states of CFT.
This conjecture is verified in many cases, so we have no doubt that
it is true.

The entire function $\CQ (\la)$ can be presented as
infinite product with respect to its zeros:
\debut \CQ (\la) =
\CQ (0)\prod\limits _{j=1}^{\infty}\(1-\({\la\over\sigma _j }\)^{2}\)
\label{prodQ}\fin
The zeros $\sigma _j$ are subject to the Bethe Ansatz equations
\debut
{\CQ(q\sigma _j) \over \CQ (q^{-1}\sigma _j)}=-1
\label{bethe}
\fin
In the next section we shall consider the solutions
to these equations in the quasi-classical limit.

As it has been said there must be another solution
($\tilde{\CQ}(\la)$)
to
the equations (\ref{q+q}).
For any two solutions to (\ref{q+q})
the ``quantum Wronskian''
$$W(\CQ ,\tilde{\CQ})(\la)=\CQ(\la)\tilde{\CQ}(\la q)-
\CQ(\la q)\tilde{\CQ}(\la )$$
is a quasi-constant:
$$W(\CQ ,\tilde{\CQ})(q\la)-  W(\CQ ,\tilde{\CQ})(\la)=0 $$
So, to find the second solution we have to solve a first-order
difference equation. Namely, let us put $W(\CQ ,\tilde{\CQ})(\la)=1$
then $\tilde{\CQ}(\la)=\CQ (\la)F(\la)$
where $F$ satisfies the equation
\debut
F(q\la)-F(\la)={1\over \CQ(\la)\CQ(q\la)}
\label{eqF}
\fin
The solution to this equation can be always found but
generally it is not a single-valued function of $\la ^2$.
Consider for simplicity the reflectionless case
$\hbar ={1\over\nu}$, with integer $\nu$. It this
case $\tilde{\CQ}$ is a single-valued function of
$\la ^2$ described as follows. Take some polynomial $P(a)$
of degree $n$ then
\debut
F(\la)=P(\la ^{2\nu})\({1\over2\pi i}\int\limits _{C_1}
{d\mu ^2\over \CQ(\mu)\CQ(\mu q)P(\mu ^{2\nu})(\mu ^2-\la ^2)}
+{1\over 2\pi i}\int\limits _{C_2}
{d\mu ^2\over P(\mu ^{2\nu})}G(\la ,\mu)\)
\label{F}
\fin
where $C_1$ encloses zeros of $\CQ (\mu)$ and
$C_2$ encloses zeros of $P(\mu ^{2\nu})$, $G(\la ,\mu)$
is the following function:
$$G(\la ,\mu)=\sum\limits _{j=1}^{\nu}\sum\limits _{l=1}^{j}
{1\over (\la^2 q^{2j}-\mu ^2)\CQ (\mu ^{-l})\CQ (\mu ^{-l+1})}
$$
The polynomial $P$ is introduced for convergence, we shall not
go into more details here.
The definition of $F$ depends on $P$, but one easily shows
that solutions with different $P$ differ by a quasi-constant.

\section{Quasi-classical wave functions.}

Our nearest goal is to understand the quasi-classical behaviour
of $\CQ (\la )$.
Consider the equations:
\debut
{\CQ(q\la ) \over \CQ (q^{-1}\la)}=-1
\label{bet}
\fin
In the case which we shall consider in this paper all the
solutions to these equations are such that $\la ^2$ is real.
A part of solutions coincides with zeros of $\CQ (\la) $ but,
obviously there are other solutions which provide zeros of $t(\la )$.
Different solutions are counted as follows:
$$\log \({\CQ(q\la _k) \over \CQ (q^{-1}\la _k)}\)=(2k+1)\pi i$$
where $-\infty <k<\infty$. We have to share these solutions between
$t (\la)$ and $\CQ (\la)$. For the quantization of periodical
analogues of soliton solutions we do it as follows.
For $-\infty<k<0$ $\la _k $ are zeros of $t (\la)$, for
$0<k<\infty$ they are zeros of $\CQ(\la)$ except for
finitely many $N_1,\cdots ,N_n >0$ which correspond to positive zeros of
$t(\la)$. We have taken the branch of logarithm such that the border
between zeros of $t$ and zeros of $\CQ$ lies at $k=0$.

In quasi-classical limit
$\hbar \to 0 $, $N_j=O(\hbar ^{-1})$ and
the zeros of $\CQ (\la)$
condense in the $\la ^2$-plane forming the cuts \cite{smr} . From comparing
with classical picture it is clear that these cuts must
coincide with the intervals: $I_1=[0,\la _1^2]$, $I_2=[\la _2^2 , \la _3^2]$,
$\cdots$ , $I_{n+1}=[\la _{2n}^2 ,\infty ]$ where $\la _j ^2$ are the branch
points defining the classical solution. There is one zero of $t (\la)$
in every interval between $I _j$ and $I _{j+1}$.
The quasi-momentum $\log \La (\la)$ can
be considered as single-valued function on the plane with these cuts.
Comparing
classical formula (\ref{simf}) 
and quantum formula (\ref{psi}) one easily finds that when $q\to 1$
$$\log \CQ  (\la)={1\over i\pi\hbar}\int\limits ^{\la}\log \La (\sigma)
{d\sigma\over\sigma}
+ O(\hbar ^0)$$
This formula must be understood as very approximate one, $\log \La (\la)$
is not a single-valued function, and its branch has to
be taken differently for different $\ga _j$.
This "tree approximation" of $\CQ (\la)$ can be
rewritten in the form:
\debut
{1\over i\pi\hbar}\int\limits _0^{\la}\log \La (\la){d\la\over\la}
={1\over \hbar}
\sum\limits _{j=1}^{n+1}\int\limits _{I_j}\log \(1-\({\la\over\sigma}\)^2 \)
\rho (\sigma)d\sigma
\label{logs}
\fin
where inside every $I_j$ the density
$$\rho (\sigma)={1\over \sigma}{\rm Re}(\log \La (\sigma)),$$
notice that $\rho (\sigma)>0$ inside $I_j$, and it vanishes as
$\sqrt{|\sigma-\la_j|} $ at the ends of intervals.
The logarithms in (\ref{logs}) have cuts in $\sigma ^2$ plane from
$\la ^2$ to $\infty$. Obviously the formula (\ref{logs}) originates from
the classical limit of the infinite product (\ref{prodQ}), and
$\rho(\sigma)$ describes the density of distribution of zeros.

We need the quasi-classical correction to $\CQ (\la)$. To find it
we have to investigate carefully the classical
limit of Bethe Anzatz equations (\ref{bethe}). It is done in
the Appendix using certain version of Destry-de Vega equations
\cite{ddv}, here we present the final result of these calculations:
\debut
\CQ (\la )=\la ^{-{1\over 2}}P^{-{1\over 4}}(\la )\ \Gamma _{\Sigma} (\la )
\exp \(
{1\over i\pi\hbar}\int\limits ^{\la}\log \La (\sigma)
{d\sigma\over\sigma}\) \label{qcQ}
\fin
where $\Gamma _{\Sigma} (\la )$ is an analogue of gamma-function related
to the surface $\Sigma $ which is defined as follows.

Let us introduce the normalized
third kind differential $\rho _{\mu}(\la) =G_{\mu}(\la )d\la $
\debut
\rho _{\mu}(\la)\sim {d\la \over (\la -\mu)}\quad
\la\sim\mu ,\qquad
\int _{b_j}\rho _{\mu}(\la)  =0\qquad \forall j\non
\fin
Then
\debut
d\log \Gamma _{\Sigma}(\la)=
-\lim _{N\to\infty}\(
{1\over2\pi i}\int\limits _{C_N}\rho _{\mu}(\la) d\log (\La (\mu) ^2 -1)
-{L\over \pi i}\log {\pi i N \over L} \rho (\la)
\)
\label{gamma}
\fin
where the integral is taken over $\mu$, and the contour $C_N$ encloses
the points $-i\mu _j$ with $1\le j\le N$,
$$\rho (\la)={\rm lim}_{\mu\to\infty}\mu \rho _{\mu}(\la)  $$
The function $\Gamma _{\Sigma }(\la)$ is not single-valued
on $\Sigma $, but it is single-valued on the plane with cuts (fig.3).

The analogy with usual $\Gamma$-function is obvious. If
we take, in particular, the vacuum state the corresponding
Riemann surface $\Sigma _{vac}$ is Riemann sphere and
$$\Gamma _{\Sigma _{vac}}(\la )=\la
\({\pi i\over L}\)^{L\la\over\pi i}\Gamma \({L\la\over \pi i}\)$$

The function $\Gamma _{\Sigma}$ satisfies the important functional
equation:
\debut
 \Gamma _{\Sigma}(\la)\Gamma _{\Sigma}(-\la)=
{2\pi ^2\la\sqrt{P(\la )}\over \Lambda(\la)-\Lambda ^{-1}(\la)}
\label{gaga}
\fin

Before going further into study of the wave functions
let us consider the quasi-classical analogue of the quasi-momentum.
By definition
\debut \Lambda _q(\la)={\CQ(\la q^{1\over 2})
\over\CQ(\la q^{-{1\over 2}})},\label{Q/Q}\fin
its asymptotics are (\ref{ass}):
\debut
\log\Lambda _q(\la)
\sim \la ^p L+\sum\limits _{k\ge 1}\la ^{-(2k+1)p}I_{2k-1}\non
\fin
recall that $p={1\over 1-\hbar}$.
Obviously one has quasi-classically:
\debut
\log\Lambda _q(\la)\sim _{\hbar\to 0}\log\La (\la)
+{\hbar}\(\log \la \ \la{d\over d\la}\log\La (\la)+
\sum\limits _{k\ge 1}\la ^{-(2k+1)p}\delta I_{2k-1}\)
\label{y}\fin
where $\delta I_{2k-1}$ are quasi-classical corrections to the
integrals of motion. The non-trivial term of this formula is that
containing $\log \la$. Let us show that our previous formulae
agree with this asymptotic behaviour.
From (\ref{qcQ}) one finds:
\debut
\log\Lambda _q(\la)=\log\La (\la)+i\hbar\la{d\over d\la}
\log(\la ^{-{1\over 2}} P(\la)^{-{1\over4}}
\Gamma _{\Sigma}(\la))+O(\hbar ^2)
\label{x}
\fin
To compare the formulae (\ref{x}) and (\ref{y})
we need to know the asymptotics of
$\Gamma _{\Sigma}$. It is quite clear from the functional equation
(\ref{gaga}) that asymptotically
$$\log(\la^{-{1\over 2}}
P(\la )^{-{1\over 4}}\Gamma _{\Sigma}(\la))={1\over \pi i}\log\la \ \log \La(\la)+
\sum\limits _{k=1}^{\infty} c _k\la ^{-2k+1}
$$
where $c_k$ are certain coefficients.
Now we see that the formula (\ref{x}) has indeed the
same structure as (\ref{y}). Actually, the coefficient
$c_k$ can be
evaluated explicitly providing
quasi-classical corrections to the integrals of motion.

Finally, we have to require that $\CQ (\la )$ is single-valued on the
plane of $\la ^2$ with cuts (fig.3). This requirement leads to
the Bohr-Sommerfeld quantization conditions:
\debut\int\limits _{a_j}d\log \CQ (\la)=2 \pi i N_j
\label{bohr}
\fin
obviously $N_j$ is the number of zeros of exact quantum $\CQ$ in
corresponding interval,
in the quasi-classical region $N_j$ 
are of order $\hbar ^{-1}$. These quantization conditions are
implicitly the quantization conditions on the moduli
$\tau _1, \cdots ,\tau _n $.

The following important circumstance must be emphasized. The
function $\CQ$ constructed in \cite{blz}
is defined in $\la ^2$ plane, so, it has as its
quasi-classical limit the function constructed above defined in
$\la ^2$ plane or, equivalently, in the upper $\la$ plane.
However in our construction this function allows
analytical continuation to the lower half-plane (second sheet).
This analytical continuation is related to the second solution of
the Baxter equation $\tilde{\CQ}$ discussed in the previous section.

Let us construct the quasi-classical wave function.
We had the formula
$$\Psi =\prod\limits _{j=-\infty}^{n-1}Q_j(\ga _j)$$
As it has been said
the functions $Q_j(\la)$ differ from $\CQ(\la)$ by
certain degree of $\la$.
Let us explain the origin of this difference. First,
notice that the expression for $\CQ$ (\ref{qcQ}) is not
uniquely defined on $\Sigma$ because $d\log \La$ has non-zero
$b$-periods (all of these periods are equal to $2\pi i$,
so adding them we would multiply $\CQ(\la)$ by $\la ^{2 k\over \hbar}$).
Let us fix the branch of $\log \La(\la)$ as follows:
make cuts over the cycles $c_j$ (fig.4) and require
$\log \La(0)=0$ at infinity.
Now if we understand the integral in (\ref{qcQ}) as
$$\int\limits ^{\la}p(\sigma){d\sigma\over\sigma}=
\int\limits  _{0}^{\la}p(\sigma){d\sigma\over\sigma} $$
The functions $Q_j(\la)$ are defined as
\debut
Q_j(\la )=\la ^{ j\over\hbar}\CQ(\la)
\label{Qj}
\fin
This prescription is chosen for the following reasons.
\newline
1. For $j<0$ it ensures the existence of saddle point
of $Q_j$ at $i\mu _j$, the point where $\ga _j$ is
situated in classics.
\newline
2. For $j>0$ it makes the action to satisfy proper reality
condition along a classical trajectory.
The problems of reality are discussed in details in \cite{bbs}.

The prescription for quantization of cKdV explained in \cite{bbs}
consists in taking the analytical continuation of rKdV. In
particular it corresponds to the following rule of hermitian
conjugation:
\debut
\Psi ^{\dag}(\ga )=\bar{\Psi (\bar{\gamma})}
\label{hermit}
\fin
which is the same as
$$
\Psi ^{\dag}(\ga )= \Psi (\gamma ^c)
$$
where $\ga ^c=\ga$ for classically confined coordinates
and $\ga ^c =-\ga $ for classically moving particles.

\section{Quasi-classical matrix elements.}

To construct the matrix elements we need to know also
the measure of integration in the space of functions of $\gamma$.
Comparing the expression for the Liouville measure (\ref{mes})
with the formula (\ref{gaga}) we assume that the measure of
integration is given quasi-classically by
$$W(\ga)=\prod\limits _{-\infty <i<j}^{n-1}(\ga _i^2 -\ga _j^2)$$

Thus the matrix elements of operators in $\ga$-representation
are given by:
\debut
\langle \Psi |O|\Psi '\rangle ={1\over \CN \CN '}
\prod\limits _{j=-\infty}^{n-1}\int\limits _
{-\infty}^{\infty}d\ga _j \ W(\ga)
\Psi (\gamma ^c)O(\ga)
\Psi '(\gamma )
\label{me}
\fin
where $\CN $, $\CN '$ are norms of the wave functions,
$O(\ga )$ is the operator $O$ in $\ga$-representation.
Let us first consider the norms.

Take the wave function for the periodical $n$-soliton solution $\Psi(\ga)$
and consider
$$ \CN ^2=
\prod\limits _{j=-\infty}^{n-1}\int\limits _
{-\infty}^{\infty}d\ga _j \ W(\ga)
\Psi (\gamma ^c)
\Psi (\gamma )$$
Consider first the integral over $\ga _{-j}$.
It is of form
$$\int\limits _
{-\infty}^{\infty} F(\ga )\exp\({2\over i\pi \hbar}\int\limits ^{\ga }
(\log \La (\sigma ) -j\pi i){d\sigma\over\sigma}\)d\ga
$$
where $F(\la)$ is finite when $\hbar\to 0$.
So, it is sitting on the stationary point $i\mu _j$ (recall that
$\log \La (\mu _j)=j\pi i$).

Now consider the integral with respect to $\ga _j$ ($j>0$).
By definition of branch of $\log \La (\la)$ one sees that on the
real axis
$$\log \La (\la)+\log \La (-\la)=\pi ik\qquad {\rm for}\quad
|\la _{k}|<|\la |<|\la _{k+1}|$$
where we put $\la _0=0,\ \la _{2n+1}=\infty$.
Together with (\ref{gaga}) it gives
$$\CQ(\ga)\CQ(-\ga)\ga ^{{2 j\over\hbar}}=
{1\over \La(\ga)-\La ^{-1}(\ga)}
\prod\limits _{k=1}^{2j}\la _k^{{1 \over\hbar}} $$
for $|\la _{2j}|<|\la |<|\la _{2j+1}|$ and exponentially 
(with ${1\over\hbar}$ in exponent) smaller
everywhere else. So, the integral with respect to $\ga _j$ is
sitting on these two segment of real axis or, putting it differently,
on the cycle $a_j$.

After these expansion let us present the final result of
calculations.
We denote by $\La _q$ the quasi-classical approximation given by (\ref{x}).
\debut
&\CN ^2 =
\Phi (\La _q)
\prod\limits _{k=1}^{2n}\la _k^{{ k \over\hbar}}
\int\limits _{a _0}d\ga _0
\cdots\int\limits _{a _{n-1}}d\ga _{n-1}\prod\limits _{j=0}^{n-1}
{1\over \La(\ga _j)-\La ^{-1}(\ga _j)}
\prod\limits _{k=1}^{\infty} \(1-{\ga ^2 _j\over\mu _{-k}^2}\)
\prod\limits _{i<j}(\ga _i^2-\ga _j^2)
\non\\
&
= \Phi (\La _q)
\prod\limits _{k=1}^{2n}\la _k^{{ k \over\hbar}}\ \Delta
\label{normm}
\fin
where
$$\Delta=
\int\limits _{a _0}d\ga _0
\cdots\int\limits _{a _{n-1}}d\ga _{n-1}\prod\limits _j
{1\over \sqrt{P(\ga)}}\prod\limits _{i<j}(\ga _i^2-\ga _j^2),
$$
and
\debut
&\Phi (\La _q)=
\exp \( {1\over 2\pi i}\int\limits _C \log (\La _q^2(\mu)-1)\[
{2\over i\pi\hbar}\({d\over d\mu}\log\La _q(\mu)\) \log\mu+
{1\over 2}{d\over d\mu}\log \({d\over d\mu}\log\La _q(\mu)\)\]d\mu-
\right.\non\\
&\left.-
{1\over \pi ^2}\int\limits _C\int\limits _C
\log (\La _q^2(\mu _1)-1)\log (\La _q^2(\mu _2)-1){\mu _1\mu _2\over
(\mu _1^2-\mu _2 ^2)^2}d\mu _1 d\mu _2
\),\label{Phi}
\fin
the contour $C$ encloses the points $i\mu _{-k}$.
The formula (\ref{normm}) must be interpreted as follows:
$\Delta $ gives the volume of the coordinate space
of the finite-zone solution while the rest describes the measure in
orthogonal, momentum, direction.
Actually the integrals in (\ref{Phi}) are divergent, one has to
divide by the norm of vacuum  to make them finite. This
divergence does not affect our further calculations, so, we shall
ignore it.

Let us consider now the matrix elements. We shall take the simplest
operator $T(0)$ which is classically the same as $u(0)$.
On a classical finite-zone solution
$u=\sum\ga _j^2-{1\over 2}\sum \la _j ^2$. The prescription for
the symbol of this operator in quasi-classical approximation
is:
$$T(0)=\sum \ga _j^2-{1\over 4}\sum \(\la _j^2 +(\la _j ')^2\)$$
where $\la$, $\la '$ are branch points corresponding to classical solutions.
We take this symmetric prescription because it looks the most
natural and gives correct answer in $L\to\infty$ limit \cite{bbs}.
The calculation of the matrix element is similar to the calculation
of norm. The only important point to realize is that the
stationary points move to those solving the equation
$$\La (\mu)\La '(\mu)=1$$
The final result is
\debut
&\langle \Psi |T(0)|\Psi '\rangle ={1\over\sqrt{\Delta\Delta '}}
\ {\Phi (\sqrt{\La _q\La '_q})\over
\sqrt{\Phi (\La _q)
\Phi (\La '_q)}}\non\\
&\times\int\limits _{-\infty}^{\infty}d\ga _0\cdots
\int\limits _{-\infty}^{\infty}d\ga _{n-1}\prod\limits _j
\CQ(-\ga _j)\CQ '(\ga _j)
\exp \({1\over\pi i}\int\limits _C{\mu\over\ga _j^2-\mu ^2}
\log (\La (\mu)\La '(\mu) -1)d\mu\)\non\\&\times
\prod\limits _j \ga _j ^{{2 j\over\hbar}}
\prod\limits _{i<j}(\ga _i^2-\ga _j^2)
\(\sum \ga _j^2-{1\over 4}\sum (\la _j^2+(\la _j')^2)\)
\label{pff}
\fin
where the contour $C$ encloses the zeros of $\La (\mu)\La '(\mu)-1$
lying on positive imaginary half-axis. 
Notice that the exponential growth at infinity is cancelled in the
combination $\CQ(-\ga _j)\CQ '(\ga _j)$.
Before discussing further this
formula let us show that it gives correct result in the limit
$L\to\infty$.

In the classical case when $L\to\infty$ the Riemann surface
degenerates: with exponential in $L$ precision $\la _{2j-1}\rightarrow
\tau _j\leftarrow\la _{2j}$. The quasi-momentum becomes an elementary
function:
\debut
\La (\la)=
e^{L\la}
\prod\limits _{j=1}^n\({\tau _j-\la\over\tau _j +\la}\)
\(1+O(e^{-\la _1 L})\)
\label{Lacl}
\fin
In the quantum case one has \cite{fst}:
\debut
\La _q (\la)=
e^{L\la^p}
\prod\limits _{j=1}^n\({\tau _j^p-\la^p\over\tau _j ^p+\la ^p}\)
\(1+O(e^{-\la _1^p L})\)
\label{Laqinf}
\fin
Let us first check that our quasi-classical formula for $\CQ$ agrees
with this result.
Consider the formula (\ref{gamma}). On the degenerate surface the
differential $\rho $ turns into
\debut
\rho _{\mu}(\la)=\({1\over(\la -\mu )} +o(L^{-1})\)d\la
\non\fin
One finds that when $L\to\infty$
\debut
&d\log \Gamma _{\Sigma}(\la)=
-\lim _{N\to\infty}\(
{1\over2\pi i}\int\limits _{C_N}\rho _{\mu}(\la)d\log (\La (\mu) ^2 -1)
-{1\over \pi i}\log {\pi i N \over L}\rho (\la )
\)\to\non\\&
\lim _{M\to\infty}
{1\over \pi i}\(\int\limits _{-iM}^0
{1\over \la -\mu}d\log \La (\mu)-
L\log M \)d\la \label{p}
\fin
where we have integrated by parts and took into account that $\La (\la)$
is exponentially big (small) in the right (left) half plane.
Calculating the integral in (\ref{p}) one gets:
$$d\(\la ^{-{1\over 2}}
P(\la )^{-{1\over 4}}\log \Ga _{\Sigma}(\la )\)\to{1\over \pi i}\(
\log \la\ d\log \La (\la) +\sum {2\tau _j\log \tau _j\over \tau _j^2-\la ^2}
d\la \)
$$
Now we substitute this expression into the definition (\ref{Q/Q}).
Obviously,
$$ \la \hbar \(
\log \la\ {d\over d\la }
\log \La (\la) +\sum {2\tau _j\log \tau _j\over \tau _j^2-\la ^2}\)
=\log \La _q(\la)-\log \La (\la) +O(\hbar ^2)$$
which proves the consistency of our quasi-classical formulae with exact
quantum formula (\ref{Laqinf}).

The calculation of the matrix element in
$L\to\infty$ limit is straightforward, but bulky.
The main simplification is due to the fact that the integrals
containing $\log (\La ^2(\mu)-1)$ or $\log (\La (\mu)\La '(\mu)-1)$
can be
evaluated as it has been done in the calculation of $d\log
\Ga _{\Sigma}(\la)$. The only non-trivial part of the calculation
is that concerning $\Delta$. Recall that the differential $d\log \La $
is a normalized second kind differential on the surface.
Requiring that $\La $ is given by (\ref{Lacl}) in $L\to\infty$
limit we actually fix completely the rule of degeneration of the
surface in this limit. In particular, one easily finds the limiting
values of the normalized holomorphic differentials, and shows that
$$\Delta\to {1\over\prod\limits _{i<j}(\tau _i^2 -\tau _j^2)}\ L^n $$
The appearance of $L^n$ is not surprising because in $L\to\infty$
limit the eigenstates are normilized with $\delta$-functions.

To make easier the comparison of the final result with
known exact formulae \cite{book} it is convenient to
introduce usual rapidity notations:
$$\beta _j={1\over p}\log\tau _j ,\qquad \al _j ={1\over p}\log\ga _j$$
Then the final result of the calculation is:
\debut
&\langle \Psi |T(0)|\Psi '\rangle =L^n\ P\ P'
\ \prod\limits _{i<j}\zeta (\be _i-\be _j)
\prod\limits _{i<j}\zeta (\be _i'-\be _j')
\prod\limits _{i,j}\zeta (\be _i-\be '_j -\pi i)\non\\&\times
\int \al _1\cdots\int d\al _n \prod\limits _{i=0}^{n-1}
\prod\limits _{j=0}^{n}\varphi (\al _i-\be _j-{\pi i\over 2})
\prod\limits _{j=0}^{n}\varphi (\al _i-\be '_j+{\pi i\over 2})
\non\\&\times
\prod\limits _{j}e^{{\pi\over\xi}(2j-n+1)\al _j}
\prod\limits _{i<j}\sinh (\al _i-\al _j)
\(\sum e^{\al _j}-{1\over 2}\sum e^{\be _j}-{1\over 2}\sum e^{\be '_j}\)
\label{book}
\fin
where
$$\int =\int\limits _{-\infty}^{\infty}-
\int\limits _{-\infty-\pi i}^{\infty-\pi i},$$
the functions $\zeta $ and $\varphi$ are given (up to some constants) by
\debut
&\varphi (\be )={1\over\sqrt{\cosh \be}}\exp\(-{2\over\xi}
\int\limits _0^{\infty}{\sinh ^2 {k\be\over 2}
\over k^2\cosh {k\pi\over 2}}dk\)\non\\
&\zeta (\be )=\sqrt{\sinh {k\be\over 2}}\exp\({1\over\xi}
\int\limits _0^{\infty}{\sinh ^2 {k\over 2}(\be+\pi i)
\over k^2\cosh ^2{k\pi\over 2}}dk\)
\non
\fin
which is exactly the quasi-classical limit of corresponding
functions used in \cite{book}.
Finally,
\debut
P=\exp \({\pi\over\xi}
\sum (2j-n-1)\be _j\)\prod\limits _{i<j}
\sqrt{\sinh (\be _i-\be _j)}\ \varphi(\be _i-\be _j +{\pi i\over 2})
\non\fin
and similarly for $P'$.
Notice that $|P|=1$.

One can show that the formula (\ref{book}) basically coincides with the
quasi-classical limit of modest modification
(similar to one done in \cite{bbs}) of usual form factors
formulae \cite{book}   the only difference being due to the phases $P$,
$P'$. This difference is quite understandable: in quasi-classical construction
we get automatically the states which are symmetric with
respect to permutation of particles, while in usual form factor formulae
the states are used which produce S-matrices under these permutations.
This difference in normalization is responsible for presence of the phases.
Notice that the presence of the functions $\zeta$ in the quasi-classical
result is due to the contribution of ``vacuum'' particles. This contribution
could not be reproduced in more naive approach of the paper \cite{bbs}.

It is clear that we are not very far from the exact quantum formula
for the form factors in finite volume.
The main feature of both quasi-classical formula  (\ref{pff})
and the exact formula in infinite volume \cite{book} is
that they are given by products of certain
multiplier which is independent on particular local
operator and finite-dimensional integral depending on the local
operator. Will this structure will hold for the exact formula
in finite volume? This is not clear, but this is the only chance
for the formula to be efficient.

\section{Appendix.}
In this Appendix we give details of calculation of the quasi-classical
limit of $\CQ $. Recall that
$$\CQ (\la)=\CQ (0)\prod\limits _{j=1}^{\infty}
\(1-\({\la\over\sigma _j}\)^2\)$$
where the zeros $\sigma _j ^2$ are real positive.
Following \cite{ddv} introduce
$$a(\la)={\CQ (\la q)\over \CQ (\la q^{-1})}$$
Using the fundamental equation
$$t(\la)\CQ (\la)=\CQ (\la q) +\CQ (\la q^{-1})  $$
one finds:
\debut
\log a(\la) -\log a(\la q^{-1}) =
-{1\over2\pi i}\int\limits _C\log (a(\mu)+1)d\log \({\la ^2q^2 -\mu ^2\over
\la ^2q^{-2} -\mu ^2}\)\label{dv}\fin
where the contour $C$ encloses the points $\sigma _j^2$ and $q^2 \sigma _j^2$
in the plane of $\mu ^2$.

We know that in the quasi-classical limit the zeros
$\sigma _j$ are dense inside $n+1$ intervals corresponding to
cuts on the fig.3. Let us present quasi-classically
$$a(\la)=a_0(\la )\(1+\hbar\pi x(\la ) +O (\hbar ^2)\)$$
In the order $\hbar ^1$ the equations (\ref{dv}) give
\debut
{d\over d\la ^2}\log a_0(\la)=-{1\over \pi i}\int _C\log
(a_0 (\mu )+1){1\over (\la ^2 -\mu ^2)^2}d\mu ^2
\label{dv0}
\fin
where the contour $C$ encloses cuts on fig.3. We know in advance from
classical consideration that $a_0(\la)=\La ^2(\la)$. Let us check, first,
that this $a_0(\la)$ indeed satisfies the equation (\ref{dv0}).
We have
$$  \log (\La ^2(\mu)+1)={1\over 2}\log \La (\mu)+\log T(\mu)$$
where $T=\La +\La ^{-1}$ is entire function without zeros
on the cuts. So, the integrals in (\ref{dv0})  with $\log T(\mu) $
disappear leaving an obvious identity. It is clear, on the other
hand, that this is the only way to satisfy this equation.
Notice that the equation does not have solution for arbitrary
positions of the branch points $\la _j^2$: the equations (\ref{period})
have to be satisfied.

Consider now the equation (\ref{dv}) in the order $\hbar ^2$.
After some simple transformations it can be rewritten as
\debut
{1-\La ^2(\la )\over 1+\La ^2(\la )}
\la ^{-1}x (\la)=i{d\over d\la }\log \La (\la)+{1\over \pi i}
\int\limits _ {C _+}
{1-\La ^2(\mu )\over 1+\La ^2(\mu )}
x(\mu){\la\over\mu^2-\la^2}{d\mu\over \mu}
\label{dv1}
\fin
where
and the contour $C _+$ encloses all zeros on $\La ^2 +1$ (or $T$)
in the $\mu ^2$
plane.
This equation shows that $x(\la )$ allows analytical continuation to
the surface $\Sigma $ which is realized, as usual, as the plane
with cuts (fig.4). Consider the differential
$$\varphi (\la)={1-\La ^2(\la )\over 1+\La ^2(\la )}
x (\la) {d\la\over \la} $$
The equation (\ref{dv1}) shows that\newline
1. $\ \varphi (\la)$ is holomorphic differential on the surface
with infinitely many simple poles at the points $\pm\tau _j$
(zeros of $\La ^2+1$).
\newline
2. Since the function $x (\la)$ is regular on the first sheet
of $\Sigma $ (in the upper half-plane) $\varphi (\la)$ has
simple zeros at the points $i\mu _j$ (zeros of $\La ^2-1$).
\newline
3. The equation holds:
\debut
\varphi (\la)+\varphi (-\la)= 2id\log \La (\la)
\label{p+p}
\fin
\newline
4. The $a$-periods of $\varphi $ vanish.

Let us show that these four conditions are sufficient to
satisfy the equation (\ref{dv1}).
Consider the canonical symmetric second kind differential $\omega (\la ,\mu)=
G(\la ,\mu )d\la d\mu $:
$$\omega (\la ,\mu)=\omega (\mu ,\mu),\qquad
\omega (\la ,\mu)\sim {d\la d\mu\over (\la -\mu)^2}\quad \la\sim\mu ,
\qquad \int _{a _j}G(\la ,\mu )d\la =0\quad\forall j
$$
It is clear that on our hyper-elliptic surface
\debut
G(\la ,\mu ) +G(\la , -\mu )  = {\la\mu\over (\la ^2-\mu ^2)^2}
\label{G+G}
\fin

Let us show that the following
equation holds for $\varphi$ satisfying the conditions 1-4:
\debut
\varphi (\la)=i{d\over d\la }\log \La (\la)+{1\over 2\pi i}
\int\limits _ {C _+ +C_-}
\varphi (\mu )
\int\limits ^{\mu}\omega (\la ,\mu)
\label{dv2}
\fin
where $C _+$  ($C _-$ ) enclose zeros on $\La ^2+1$ in upper (lower)
half-planes. Indeed, the contour $C_+ +C_-$ encloses all the
singularities of the integrand except for that at $\mu =\la$ and
that at $\mu =\infty$. Obviously residue of the integrand at $\mu =\la$
equals $\varphi (\la)$. Thus using the Riemamm bilinear relation
one has
\debut &\int\limits _ {C _+ +C_-}
\varphi (\mu )
\int\limits ^{\mu}\omega (\la ,\mu) =2\pi i \varphi (\la )+
\lim _{R\to\infty}\int\limits _ {S _+ +S_-}
\varphi (\mu )
\int\limits ^{\mu}\omega (\la ,\mu) +\non\\&+
\sum\limits _{i=1}^n \(
\int\limits _{a_i}  \omega (\la ,\mu)
\int\limits _{b_i}\varphi (\mu )  -
\int\limits _{a_i}\varphi (\mu )
\int\limits _{b_i}\omega (\la ,\mu) \)
\non
\fin
The latter sum vanishes because the differential $\omega$ and
$\varphi $ have vanishing $a$-periods. The contours $S _+$ ($S_-$)
are half-circles in the left (right) half-planes of radius $R$.
Using the condition 3 one has
$$ \lim _{R\to\infty}\int\limits _ {S _+ +S_-}
\varphi (\mu )
\int\limits ^{\mu}\omega (\la ,\mu) =
\lim _{R\to\infty}2\int\limits _ {S _+}
id\log \La (\mu )\int\limits ^{\mu}\omega (\la ,\mu) =
2\pi L\lim _{\mu\to\infty} \mu ^2{\omega (\la ,\mu)\over d\mu}$$
Recall mow that $\omega (\la ,\mu)$ is normalized second kind
differential with double pole at $\la =\mu$ while
$d\log \La (\la )$ is a normalized second kind
differential with singularity at infinity where it behaves
as $Ld\la $. Hence
$$\lim _{\mu\to\infty} \mu ^2{\omega (\la ,\mu)\over d\mu}
=L^{-1}d\log \La (\la )
$$
which proves the equation (\ref{dv2}).
Using (\ref{p+p})  and (\ref{G+G}) and the fact that $ d\log \La (\la ) $
does not have singularities inside $C_+$ one shows that
(\ref{dv1}) follows from (\ref{dv2}).  Thus $\varphi$ satisfying
the conditions 1-4 satisfies the equation (\ref{dv1}).

Let us return to the function $x(\la)$. From (\ref{p+p}) one gets
\debut
x (\la) +x (-\la)=-2i\la{d\over d\la}\log (\La (\la)-\La ^{-1}(\la ))
\label{x+x}
\fin
Recall that
$$ \La (\la)-\La ^{-1}(\la )=\la \sqrt{P(\la)}\prod\limits _{j=1}^{\infty}
\(1-\({\la\over\mu _{-j}}\)^2\)$$
Let us present $x(\la )$ as 
$$x(\la)=2i\la{d\over d\la}\(-{1\over 2}\log (\la \sqrt{P(\la)})+
\log \Gamma _{\Sigma}(\la)\)$$
The function $\Gamma _{\Sigma}(\la)$ must be regular in the 
upper half plane, so
the simple poles of the RHS of (\ref{x+x}) must be shared between
$x(\la )$ and $x(-\la )$ in such a way that
$ d\log \Gamma _{\Sigma}(\la)$ is a differential with simple poles as
the points $-i\mu _{-j}$ with residues $1$.
Such a differential is not uniquely defined: one can add to it
holomorphic differentials. These are zero-modes of the equation (\ref{dv1}).
The origin of these zero-modes is clear: the classical 
surface is parametrized by $n$ moduli $\tau _1, \cdots , \tau _n$,
the zero-modes correspond to variation of these parameters.
Since after all we have to impose the Bohr-Sommerfeld quantization
condithions (\ref{bohr})
one can show that the final result for $\CQ$ does not
depend on these zero-modes. 
We require that the $b$-periods of $d\log \Gamma _{\Sigma}(\la)$
vanish because with this choice the quantum rapidities of
solitons don't differ from the classical ones in $L\to\infty$
limit. 
Thus one presents $d\log \Gamma _{\Sigma}(\la)$ in the
form given by the formula (\ref{gamma}).

\end{document}